\documentstyle[multicol,aps,prl]{revtex}

\begin{document}

\newcommand{\bc}{\begin{center}}
\newcommand{\ec}{\end{center}}
\newcommand{\be}{\begin{equation}}
\newcommand{\ee}{\end{equation}}
\newcommand{\beqn}{\begin{eqnarray}}
\newcommand{\eeqn}{\end{eqnarray}}

\begin{multicols}{2}
\narrowtext
\parskip=0cm

\noindent
{\large\bf Comment on "Random Walks, Reaction-Diffusion, and
  Nonequilibrium Dynamics of Spin Chains in One-Dimensional Random
  Environments"} 
\medskip


In a recent Letter Fisher, LeDoussal and Monthus \cite{fisher} studied
various properties of Sinai's model for diffusion in a
one-dimensional random environment by a real space renormalization
group (RSRG) calculation. It is claimed that despite its approximate
character the RSRG yields asymptotically exact results. In this
comment we want to show that 1) most of their results can be derived
in a rigorous way without recurring to an approximate RG scheme, and
2) that in this way far more general results can be obtained, not
restricted to the special case (the random force model) considered in
\cite{fisher}, nor to the vicinity of the critical (unbiased) point.

First we would like to point out that all what follows is valid for a
quite general one-dimensional random walk with nearest neighbor
hopping, which is characterized by the transition probabilities
$w_{i,i\pm 1}=w(i\to i\pm 1)$ for a random walker to jump from site
$i$ to site $i\pm 1$. Of particular interest is the asymmetric case,
$w_{i,i+1}\ne w_{i+1,i}$, for which the random force model considered
in \cite{fisher} is one special example.

A {\it new} prediction of \cite{fisher} is the persistence exponent for a
single walker. We would like to stress that there is an {\it exact}
formula for the persistence probability of a single walker on a strip
of width $L$, i.e.\ the probability that a walker does not return to
its starting point on the left before it leaves the system on the
right. It is given by \cite{aperiodic}
\be
p_{\rm pr}(L)
=\left[\left(1+\sum_{i=1}^L\prod_{j=1}^i
\frac{w_{j,j-1}}{w_{j,j+1}}\right)^{-1}\right]_{\rm av}
\label{persl}
\ee
and is completely analogous to a corresponding {\it exact} formula for
the surface magnetization of random transverse Ising chains (RTIC) derived
and analyzed in \cite{rw}. It can be shown rigorously that for the
binary distribution $P(w_{i,i\pm1}) = 1/2\cdot\{\delta(w_{i,i\pm1}-\lambda)
+ \delta(w_{i,i\pm1}-\lambda^{-1})\}$ in the limit $\lambda\to0$ the
above formula transforms into the survival probability of a random
walker in a {\it homogeneous} environment with one adsorbing boundary. Thus
$p_{\rm pr}(L)\propto L^{-\theta_1}$ with the persistence exponent
$\theta_1=1/2$.


Another new prediction of \cite{fisher} is the persistence properties
of the thermally averaged position $\langle x(t)\rangle$, which is
different from the single walker persistence. It is claimed that the
probability for $\langle x(t)\rangle$ not to return to the starting
value $x(0)$ before leaving a strip of width $L$ is given by the
averaged persistence probability $\overline{p}_{\rm pr}(L)\sim
L^{-\overline{\theta}}$ with $\overline{\theta}=(3-\sqrt{5})/4 =
0.19098\dots$, related to the golden mean.  This is an astonishing
prediction and very much as in the case for the bulk magnetization of
the RTIC\cite{fisherchain} demands an alternative verification, since
the character of the RG treatment in \cite{fisher} is approximate. For
this we would like to point out that for the limiting distribution
mentioned above $\overline{p}_{\rm pr}(L)$ can be calculated from a
random walk model in a {\it homogeneous} environment \cite{walk}. It
is possible to determine $\overline{p}_{\rm pr}(L)$ {\it exactly} for
strip widths up to L=14 and it turns out that usual series
extrapolation methods yield an exponent
\be
\overline{\theta}=0.191\pm0.002\:,
\ee
in very good agreement with the prediction of \cite{fisher}. Moreover,
it is possible to define a profile interpolating between the single
walker and average persistence \cite{walk}, analogous to the
magnetization profiles in the RTIC interpolating between surface and
bulk magnetization \cite{profiles}.

Finally in \cite{fisher} the analogy between the Griffiths-McCoy phase of
RTIC's and the biased case of Sinai's model has been emphasized.
However, this analogy is only utilized asymptotically, i.e.\ for small
drift close to unbiased case that corresponds to the critical point
of the RTIC. We would like to stress that this correspondence goes
much further, holds also deep in the Griffiths-McCoy phase and allows one to
derive {\it new} results for the RTIC from the knowledge about
anomalous diffusion properties of Sinai's model with drift. So, for
instance, the dynamical exponent $z(\delta)$ (with $\delta$ being the
distance from the critical point) parameterizing the Griffiths-McCoy
singularities in the RTIC is given implicitly by the exact formula 
\cite{walk}
\be
\left[\left({J \over h}\right)^{1/z(\delta)}\right]_{\rm av}=1\;.
\label{zexp}
\ee
where $J$ and $h$ are the random bonds and fields, respectively.  Note
that for any distribution of $J$ and $h$ one obtains immediately the
result $1/z=2\delta+{\cal O}(\delta^2)$, concurring with the RG
prediction.

To conclude we have shown in this comment that many of the results of
\cite{fisher} can be derived in a rigorous way without recurring to an
approximate RG scheme and that they are not restricted to the random
force model or to the vicinity of the critical point.

\bigskip
\noindent
Ferenc Igl\'oi$^a$ and Heiko Rieger$^b$
\smallskip

{\small
$^a$ Research Institute for Solid State Physics and Optics

H-1525 Budapest, P.O.Box 49, Hungary
\smallskip

$^b$ HLRZ, Forschungszentrum J\"ulich

52425 J\"ulich, Germany

}
\bigskip
\noindent
Date: 13 May 1998


\end{multicols}

\end{document}